\documentclass[twocolumn,times]{aastex63}

\usepackage[normalem]{ulem}         
\newcommand{\be}{\begin{eqnarray}}
\newcommand{\ee}{\end{eqnarray}}

\newcommand{\Ftilde}{\widetilde{F}}
\newcommand{\DM}{{\rm DM}}

\newcommand{\Gscatt}{G_{\rm scatt}}
\usepackage[utf8]{inputenc}
\usepackage{amsmath}
\newcommand{\Lagr}{\mathcal{L}}
\newcommand{\FDMunit}{pc$^{4/3}$ km$^{-1/3}$ cm$^{-1/3}$}
\newcommand{\Funit}{pc$^{-2/3}$ km$^{-1/3}$}
\newcommand{\DMunit}{pc cm$^{-3}$}

\newcommand{\Pne}{P_{\delta n_e}}
\newcommand{\cnsq}{{\rm C_n^2}}
\newcommand{\linner}{l_{\rm i}}
\newcommand{\louter}{l_{\rm o}}

\newcommand{\DMMWh}{\DM_{\rm MW,h}}

\newcommand{\DMhalohat}{\widehat{\DM}_{\rm MW, h}}

\newcommand{\Ktau}{K_\tau}
\newcommand{\Atau}{A_\tau}
\newcommand{\re}{r_{\rm e}}
\renewcommand{\Gscatt}{G_{\rm scatt}}
\newcommand{\dsl}{d_{\rm sl}}
\newcommand{\dlo}{d_{\rm lo}}
\newcommand{\dso}{d_{\rm so}}
\newcommand{\thetad}{\theta_{\rm d}}
\newcommand{\nud}{\Delta \nu_{\rm d}}

\graphicspath{{./}{figures/}}

\begin{document}

\title{Constraining Galaxy Haloes from the Dispersion and Scattering of Fast Radio Bursts and Pulsars}

\author[0000-0002-4941-5333]{Stella Koch Ocker}
\author[0000-0002-4049-1882]{James M. Cordes}
\author[0000-0002-2878-1502]{Shami Chatterjee}
\affiliation{Department of Astronomy and Cornell Center for Astrophysics and Planetary Science, Cornell University, Ithaca, New York, 14853, USA}
\correspondingauthor{Stella Koch Ocker}
\email{sko36@cornell.edu}
\received{2021 Jan 13} \revised{2021 Feb 16} \accepted{2021 Mar 1}

\shorttitle{Galaxy Haloes and FRB Scattering}
\shortauthors{Ocker, S.K., Cordes, J.M., \& Chatterjee, S.}

\begin{abstract}
    
    Fast radio bursts (FRBs) can be scattered by ionized gas in their local environments, host galaxies, intervening galaxies along their lines-of-sight, the intergalactic medium, and the Milky Way. The relative contributions of these different media depend on their geometric configuration and the internal properties of the gas. When these relative contributions are well understood, FRB scattering is a powerful probe of density fluctuations along the line-of-sight. The precise scattering measurements for \object{FRB~121102} and \object{FRB~180916} allow us to place an upper limit on the amount of scattering contributed by the Milky Way halo to these FRBs. The scattering time $\tau\propto(\Ftilde \times \DM^2) A_\tau$, where \DM\ is the dispersion measure, $\Ftilde$ quantifies electron density variations with $\Ftilde=0$ for a smooth medium, and the dimensionless constant $A_\tau$ quantifies the difference between the mean scattering delay and the $1/e$ scattering time typically measured. A likelihood analysis of the observed scattering and halo DM constraints finds that $\Ftilde$ is at least an order of magnitude smaller in the halo than in the Galactic disk. The maximum pulse broadening from the halo is $\tau\lesssim12$ $\mu$s at 1~GHz. We compare our analysis of the Milky Way halo with other galaxy haloes by placing limits on the scattering contributions from haloes intersecting the lines-of-sight to \object{FRB~181112} and \object{FRB~191108}. Our results are consistent with haloes making  negligible or very small contributions to the scattering times of these FRBs.
    
\end{abstract}

\section{Introduction}
Fast radio bursts (FRBs) propagate from as far as $\sim$Gpc distances through their local environments, the interstellar medium (ISM) and circumgalactic medium (CGM) of their host galaxy, the intergalactic medium (IGM) and any intervening galaxies or galaxy haloes, the halo and ISM of the Milky Way, and finally through the interplanetary medium (IPM) of our Solar System before arriving at the detector. Along their journey they experience dispersion and multi-path propagation from free electrons along the line-of-sight (LoS). The dispersion measure $\DM = \int n_e dl/(1+z)$, where $n_e$ is the electron density and $z$ is the redshift. Most FRBs are extragalactic and have DMs much larger than the expected contribution from our Galaxy, with the single possible exception being the Galactic magnetar source SGR 1935$+$2154 \citep{2020arXiv200510828B, 2020arXiv200510324T}. Many studies of FRB propagation have focused on the ``DM budget," constraining the relative contributions of intervening media to the total observed FRB DM, with particular attention paid to determining the DM contribution of the IGM, which in principle can be used to estimate the distance to an FRB without a redshift \citep[e.g.,][]{2015MNRAS.451.4277D,2019ApJ...886..135P}. For FRBs with redshifts the subsequent intergalactic $\DM(z)$ relationship   can be used to measure the cosmic baryon density, as was first empirically demonstrated by \cite{2020Natur.581..391M}. Arguably the least constrained DM contribution to FRBs is that of their host galaxies, which has been estimated in only one case using Balmer line  observations \citep[][]{2017ApJ...834L...7T}.

\indent Understanding FRB propagation requires study of not just dispersion but also scattering.  Bursts propagate along multiple ray paths due to electron density fluctuations, which  leads to detectable chromatic effects like pulse broadening, scintillation, and angular broadening. These effects are respectively characterized by a temporal delay $\tau$, frequency bandwidth $\nud$, and full width at half maximum (FWHM) of the scattered image $\thetad$. Scattering effects generally reduce FRB detectability, obscure burst substructure, or produce multiple images of the burst, and may contaminate emission signatures imprinted on the signal at the source \citep[e.g.,][]{2017ApJ...842...35C, 2019ApJ...876L..23H, 2020MNRAS.497.3335D}. On the other hand, scattering effects can also be used to resolve the emission region of the source \citep{2020ApJ...899L..21S} and to constrain properties of the source's local environment \citep[for a review of FRB scattering see][]{2019A&ARv..27....4P, 2019ARA&A..57..417C}. Since the relationship between $\tau$ and DM is known for Galactic pulsars \citep[e.g.,][]{1997MNRAS.290..260R, 2004ApJ...605..759B,2015ApJ...804...23K,2016arXiv160505890C}, it is a promising basis for estimating the DM contributions of FRB host galaxies based on measurements of $\tau$ (Cordes et al., in prep). In order to use scattering measurements for these applications, we need to assess how intervening media contribute to the observed scattering (i.e., a ``scattering budget"). To disentangle any scattering effects intrinsic to the host galaxy or intervening galaxies, we need to accurately constrain the scattering contribution of the Milky Way. 

\indent Broadly speaking, an FRB will encounter ionized gas in two main structural components of the Milky Way, the Galactic disk and the halo. The Galactic disk consists of both a thin disk, which has a scale height of about 100 pc and contains the spiral arms and most of the Galaxy's star formation \citep[e.g.,][]{2002astro.ph..7156C}, and the thick disk, which has a scale height of about 1.6 kpc and is dominated by the more diffuse, warm ($T\sim10^4$ K) ionized medium \citep[e.g.,][]{2020ApJ...897..124O}. The halo gas is \added{thought to be} dominated by the hot ($T\sim 10^6$ K) ionized medium, and most of this hot gas is contained within 300 kpc of the Galactic center \added{\citep[e.g.,][]{2017ApJ...835...52F}}. While the DMs and scattering measurements of Galactic pulsars and pulsars in the Magellanic Clouds predominantly trace plasma in the thin and thick disks, extragalactic sources like FRBs are also sensitive to gas in the halo.

\indent In this paper we assess the contribution of galaxy haloes to the scattering of FRBs. We demonstrate how scattering measurements of FRBs can be interpreted in terms of the internal properties of the scattering media, and apply this formalism to galaxy haloes intervening LoS to FRBs. We first assess scattering from the Milky Way halo using two case studies: FRB 121102 and FRB 180916. \deleted{These FRBs are seen towards the Galactic anti-center and have highly precise localizations and host galaxy associations, in addition to precise scattering measurements.} \added{These FRBs have the most comprehensive, precise scattering measurements currently available, in addition to highly precise localizations and host galaxy associations}. Due to their location close to the Galactic plane, the emission from these sources samples both the outer spiral arm of the Galaxy and the Galactic thick disk, and the scattering observed for these FRBs is broadly consistent with the scattering expected from the spiral arm and disk. Only a minimal amount of scattering is allowed from the Galactic halo along these LoS, thus providing an upper limit on the halo's scattering contribution. We then extrapolate this analysis to two FRBs that pass through haloes other than those of their host galaxies and the Milky Way, FRB 181112 and FRB 191108.

\indent In Section~\ref{sec:theory} we summarize the formalism relating electron density fluctuations and the observables $\tau$, DM, $\nud$, and $\thetad$, and describe our model for the scattering contribution of the Galactic halo. A new measurement of the fluctuation parameter of the Galactic thick disk is made in Section~\ref{sec:disk} using Galactic pulsars. An overview of the scattering measurements for FRB 121102 and FRB 180916 is given in Section~\ref{sec:frbs}, including an updated constraint on the scintillation bandwidth for FRB 121102 and a comparison of the scattering predictions made by Galactic electron density models NE2001 \citep{2002astro.ph..7156C, 2003astro.ph..1598C} and YMW16 \citep{2017ApJ...835...29Y}. The FRB scattering constraints are used to place an upper limit on the fluctuation parameter of the Galactic halo in Section~\ref{sec:halo}. A brief comparison of the FRB-derived limit with scattering observed towards the Magellanic Clouds is given in Section~\ref{sec:MCs} and scattering constraints for intervening galaxy haloes are discussed in Section~\ref{sec:conc}.

\section{Modeling}\label{sec:theory}
\subsection{Electron Density Fluctuations and Scattering}

\indent We characterize the relationship between electron density fluctuations and scattering of radio emission using an ionized cloudlet model in which clumps of gas in the medium have a volume filling factor $f$, internal density fluctuations with variance $\epsilon^2 = \langle (\delta n_e)^2 \rangle/{n_e}^2$, and cloud-to-cloud variations described by $\zeta = \langle {n_e}^2 \rangle/\langle {n_e} \rangle^2$, where $n_e$ is the local, volume-averaged mean electron density \citep[][]{1991Natur.354..121C, 2002astro.ph..7156C, 2003astro.ph..1598C}.   We assume that internal fluctuations follow a power-law wavenumber spectrum of the form \citep[][]{cfrc87} 
$\Pne(q) = \cnsq q^{-\beta} \exp(-(q\linner / 2\pi)^2)$ that extends over a  wavenumber range
$2\pi / \louter \le q \lesssim 2\pi / \linner$  defined by the outer  and inner scales, $\louter, \linner$, respectively.   We adopt a wavenumber index $\beta = 11/3$ that corresponds to a Kolmogorov spectrum. Typically, $\linner \ll \louter$, but their magnitudes  depend on the physical mechanisms driving and dissipating turbulence, which  vary between different regions of the ISM.    

\indent Multipath propagation broadens pulses by a characteristic  time $\tau$  that we relate to  DM and other quantities.  For a medium with homogeneous properties,  the scattering time in Euclidean space  is $\tau({\rm DM}, \nu) = \Ktau \Atau \nu^{-4} \Ftilde \Gscatt \DM^2$ \citep[][]{2016arXiv160505890C}. The coefficient $\Ktau =  \Gamma(7/6) c^3 \re^2 / 4$, where $c$ is the speed of light and $\re$ is the classic electron radius, while the factor $\Atau \lesssim 1$ scales the mean delay to the $1/e$ delay that is typically estimated from pulse shapes.   Because $\Atau$ is medium dependent, we include it in relevant expressions symbolically rather than adopting a specific value. The scattering efficacy is determined by
 the fluctuation parameter $\Ftilde = \zeta\epsilon^2 / f (\louter^2\linner)^{1/3}$ combined with a dimensionless geometric factor,  $\Gscatt$, discussed below. 
 
\indent Evaluation  with  DM in \DMunit,  the observation frequency $\nu$ in GHz,  the outer scale in pc units,  and the inner scale in km, $\Ftilde$ has units \Funit\ and 
\be
\label{eq:taudmnu}
     \tau(\DM,\nu) \approx 48.03~{\rm ns} \,\Atau \nu^{-4} \Ftilde \Gscatt \DM^2 .
\ee
For reference,  the NE2001 model uses a similar parameter,  $F = \Ftilde l_i^{1/3}$
\citep[Eq. 11-13][]{2002astro.ph..7156C}, that relates the scattering measure SM to \DM\ and varies substantially between different model components (thin and thick disks, spiral arms, clumps). 

\indent The geometric factor $\Gscatt$ in Eq.~\ref{eq:taudmnu} depends on the location of the FRB source relative to the dominant scattering medium  and is calculated using the standard Euclidean weighting
$(s/d)(1-s/d)$  in  the integral of  $\cnsq(s)$ along the LoS.   If both the observer and source are  embedded in  a medium with homogeneous scattering strength, $\Gscatt = 1/3$, while  $\Gscatt = 1$ if the source to observer distance $d$ is much larger than the medium's thickness and either the source or the observer is embedded in the medium.    

\indent For a thin scattering layer with thickness $L$ at distance $\delta d \gg L$ from the source or observer, 
$\Gscatt \simeq 2\delta d / L \gg 1$ because of the strong leverage effect.   For thin-layer scattering of cosmological sources by, e.g., a galaxy disk or halo, $\Gscatt = 2\dsl \dlo/L\dso$ where $\dsl, \dlo$ and $\dso$ are angular diameter distances for source to scattering layer,   scattering layer to observer, and source to observer, respectively. The scattering time is also multiplied by a redshift factor $(1+ z_{\ell})^{-3}$ that takes into account time dilation and the higher frequency at which scattering occurs in the layer at redshift $z_{\ell}$, with $\DM_{\ell}$ representing the lens frame value.  We thus have for distances in Gpc and $L$ in Mpc, 
\be
\label{eq:taudmnuz}
     \tau(\DM,\nu, z) \approx 48.03~{\rm \mu s} \times
       \frac{\Atau  \Ftilde\,  \DM_{\ell}^2 }{(1 + z_{\ell})^3 \nu^4}
     	\left[\frac{2\dsl\dlo}{  L\dso} \right] .
\ee
If the layer's DM could be measured, it would be smaller by a factor $(1+z_{\ell})^{-1}$ in the observer's frame. 

The pulse broadening time is related to the scintillation bandwidth $\nud$ through the uncertainty principle $2\pi \tau \nud = C_1$, where $C_1 = 1$ for a homogeneous medium and $C_1=1.16$ for a Kolmogorov medium \citep{1998ApJ...507..846C}. Multipath propagation is also manifested as angular broadening, $\thetad$, defined as the FWHM of the scattered image of a point source. The angular and pulse broadening induced by a thin screen are related to the distance between the observer and screen, which will be discussed further in Section~\ref{sec:121102}.

\indent Measurements of $\tau$, $\nud$, and $\thetad$ can include both extragalactic and Galactic components. We use the notation $\tau_{\rm MW, d}$, $\tau_{\rm MW, h}$, and $\tau_{\rm i,h}$ to refer to scattering contributed by the Galactic disk (excluding the halo), the Galactic halo, and intervening haloes, respectively, and an equivalent notation for DM. To convert between $\nud$ and $\tau$ we adopt $C_1 = 1$. Wherever we use the notation $\tau$ and $\thetad$ we refer to the $1/e$ delay and FWHM of the autcorrelation function that are typically measured.

\subsection{Electron Density Model for the Galactic Halo}\label{sec:halomodels}
Models of the Milky Way halo based on X-ray emission and oxygen absorption lines depict a dark matter halo permeated by hot ($T\sim 10^6$ K) gas with a virial radius between 200 and 300 kpc \citep[e.g.,][]{2019MNRAS.485..648P, 2020ApJ...888..105Y, 2020MNRAS.496L.106K}. Based on these models, the average DM contribution of the hot gas halo to FRBs is about 50 \DMunit, which implies a mean electron density $n_e \sim 10^{-4}$ cm$^{-3}$. However, the DM contribution of the Milky Way halo is still not very well constrained. \cite{2020MNRAS.496L.106K} compare the range of $\DMMWh$ predicted by various halo models with the XMM-Newton soft X-ray background \citep{2013ApJ...773...92H} and find that the range of $\DMMWh$ consistent with the XMM-Newton background spans over an order of magnitude and could be as small as about 10 \DMunit. Using a sample of DMs from 83 FRBs and 371 pulsars, \cite{2020ApJ...895L..49P} place a conservative upper limit on $\DMhalohat<123$ \DMunit, with an average value of $\DMhalohat \approx 60$ \DMunit.

\indent Most models of the hot gas halo adopt a spherical density profile, but \cite{2020ApJ...888..105Y} and \cite{2020NatAs.tmp..214K} argue that a disk component with a scale height of about 2 kpc and a radial scale length of about 5 kpc should be added to the spherical halo based on the directional dependence of emission measure found in Suzaku and HaloSat X-ray observations \citep{2018ApJ...862...34N, 2020NatAs.tmp..214K}. In such a combined disk-spherical halo model, the disk would account for most of the observed X-ray emission attributed to the halo, while the diffuse, extended, spherical halo contains most of the baryonic mass. \cite{2020NatAs.tmp..214K} also suggest that significant, patchy variations may exist in the halo gas on scales $\sim400$ pc. The physical scales of the disk models fit to these recent X-ray observations are similar to the spatial scale of the warm ionized medium in the Galactic disk, and several orders of magnitude smaller than the spatial scales ($\sim 100$s of kpc) typical of spherical halo models. Whether such a disk component should really be attributed to the circumgalactic medium and not to the ISM of the Galactic disk is unclear. 

\indent \deleted{We adopt a fiducial model for the halo's density profile based on} \added{We use} the \citeauthor{2019MNRAS.485..648P} (PZ19) modified Navarro-Frenk-White (mNFW) profile \added{to model the halo density}. The mNFW profile adjusts the NFW profile's matter density cusp at the Galactic center with a more physical roll-off, giving a matter density of
\begin{equation}
    \rho(y) = \frac{\rho_0}{y^{1-\alpha}(y_0 + y)^{2+\alpha}},
\end{equation}
\deleted{where $y = K_c(r/r_{200})$}
\added{where $y = K_c\times(r/r_{200})$}, $r$ is radial distance from the Galactic center, and $r_{200}$ is the virial radius within which the average density is 200 times the cosmological critical density. The characteristic matter density $\rho_0$ is found by dividing the total mass of the halo by the volume within the virial radius. The concentration parameter $K_c$ depends on the galaxy mass; e.g., $K_c = 7.7$ for a total Milky Way halo mass $M = 1.5 \times 10^{12} M_\odot$, and can range from $K_c = 13$ for $M = 10^{10} M_\odot$ to $K_c = 5$ for $M = 10^{14} M_\odot$ for redshifts $z<0.1$ \citep{1997ApJ...490..493N}. Like PZ19, we assume that $75\%$ of the baryonic matter in a galaxy is in the halo ($f_{\rm b} = 0.75$), and the fraction of the total matter density that is baryonic is $\Omega_{\rm b}/\Omega_{\rm m} $, the ratio of the baryonic matter density to the total matter density ($\Omega_{\rm b}/\Omega_{\rm m} = 0.16$ today). If $f_{\rm b}$ is smaller, then the electron density and the predicted scattering from a halo will be smaller.

\indent The electron density profile of the halo is
\deleted{$n_e(r) = f_{\rm b}(\Omega_{\rm b}/\Omega_{\rm m})\rho(r)\frac{\mu_{\rm e}}{m_{\rm p} \mu_{\rm H}}U(r)$}

\added{\begin{equation}\label{eq:neh}
    n_e(r) \approx 0.86f_{\rm b}\times(\Omega_{\rm b}/\Omega_{\rm m})\frac{\rho(r)}{m_{\rm p}}U(r),\end{equation}}

\begin{figure}
    \centering
    \includegraphics[width=0.47\textwidth]{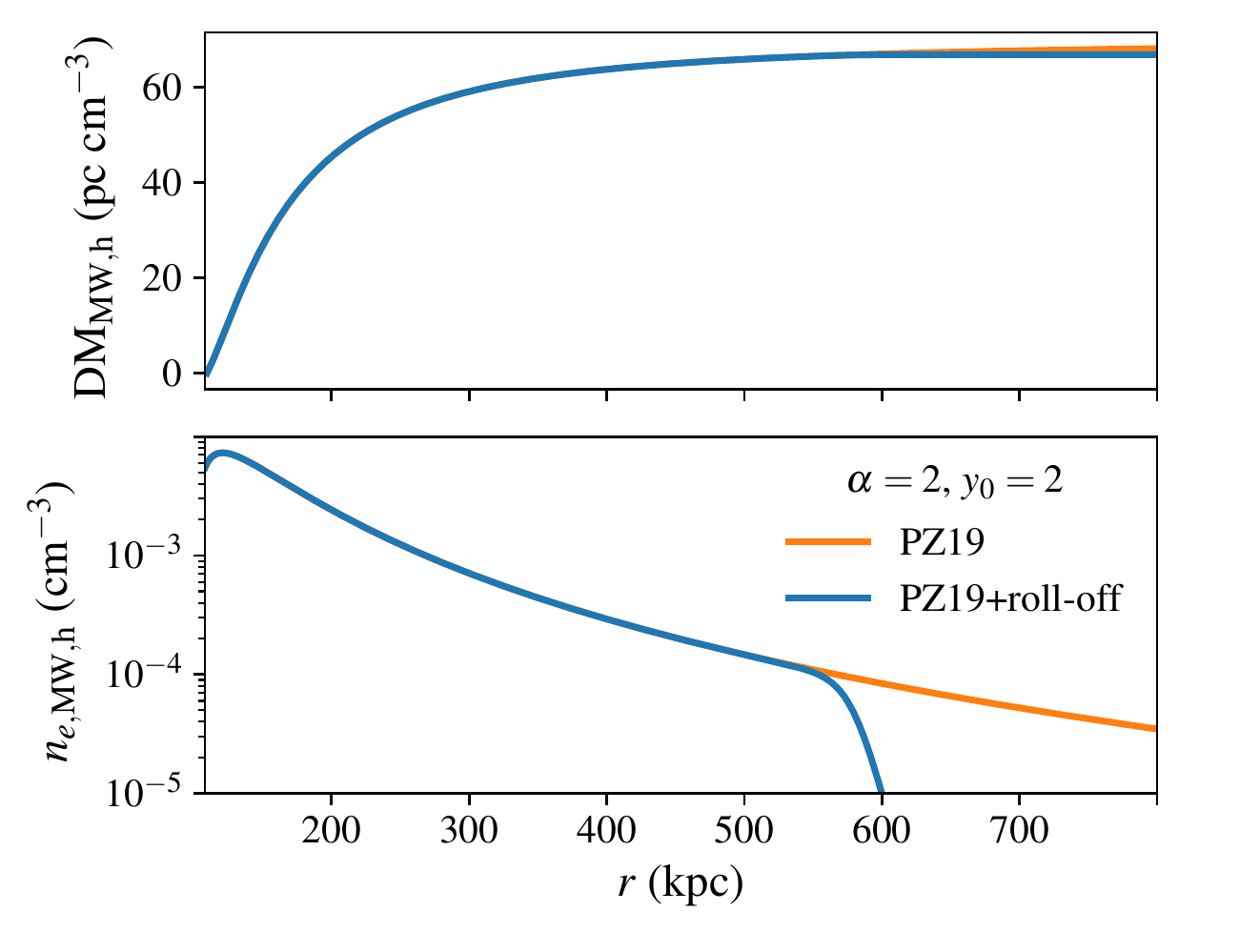}
    \caption{Electron density model and resulting DM contribution of the Milky Way halo, for an observer 8 kpc from the Galactic center. The modified NFW profile of \cite{2019MNRAS.485..648P} for $\alpha = y_0 = 2$ is shown in orange, and our \replaced{adaptation}{implementation} of the PZ19 model is shown in blue. The maximum DM contribution of the halo predicted by the model is 63 \DMunit, similar to the predictions of other halo density models for lines of sight to the Galactic anti-center \citep[e.g.,][]{2020ApJ...895L..49P, 2020ApJ...888..105Y,2020MNRAS.496L.106K}.}
    \label{fig:halomod}
\end{figure}

\deleted{where $\mu_{\rm e} = 1.12$ for fully ionized hydrogen and helium, $\mu_{\rm H} = 1.3$, and $m_{\rm p}$ is the proton mass. The function $U(r) = (1/2)\{1-{\rm tanh}[(r-r_c)/w]\}$ imposes a physical roll-off at a characteristic radius $r_c$ with a width set by $w$. Previous studies often implicitly assumed a physical boundary at the virial radius when integrating $n_e$ to obtain $\DMMWh$, despite the fact that $\rho(r)$ extends well beyond the virial radius. Nonetheless, most of the halo gas likely lies within a few times the virial radius, so we set $r_c = 2r_{200} \approx 480$ kpc and $w = 20$ kpc. A comparison of our halo density profile with the PZ19 model, evaluated for the Milky Way, is shown in Figure~\ref{fig:halomod}.}
\added{where $m_p$ is the proton mass and we have assumed a gas of fully ionized hydrogen and helium. The function $U(r) = (1/2)\{1-{\rm tanh}[(r-r_c)/w]\}$ imposes an explicit integration limit at a radius $r_c = 2r_{200}$ over a region of width $w = 20$ kpc so as to avoid sharp truncation of the model. Our implementation of the PZ19 model,
evaluated for the Milky Way, is shown in Figure~\ref{fig:halomod}.}

\subsection{Scattering from the Galactic Halo}

\indent Generally speaking, scattering from the Galactic halo traces the same plasma that gives rise to dispersion, weighted by the fluctuation parameter $\Ftilde$. To constrain $\Ftilde$ from measurements of $\tau$ and $\thetad$, we approximate the total scattering as a sum of two components: one from the disk and spiral arms of the Milky Way, which we denote with the subscript ($\rm MW,d$), and one from the Galactic halo, which we denote with the subscript ($\rm MW,h$). The total $\tau$ and $\thetad$ predicted by the model are therefore
\begin{equation}
    \tau_{\rm MW}^{\rm total} = \tau_{\rm MW,d} + \tau_{\rm MW,h}
\end{equation}
and
\begin{equation}
    \theta_{\rm d,MW}^{\rm total} = \sqrt{\theta_{\rm MW,d}^2 + \theta_{\rm MW,h}^2}.
\end{equation}
We adopt the NE2001 predictions for the $\rm MW$ components and model the halo components using Equations~\ref{eq:taudmnu} and~\ref{eq:ds}. The composite  parameter $\Atau(\Ftilde \times \DM^2)_{\rm MW,h}$ is constrained by maximizing the likelihood function
\begin{equation}\label{eq:Like}
    \Lagr((\Ftilde \times \DM^2)_{\rm MW,h} | \tau_j) = \prod_j N(\tau_{j,\rm MW}^{\rm total} - \tau_j^{\rm obs}, \sigma_{\tau,j}^2)
\end{equation}
using measurements $\tau^{\rm obs}$ and $\thetad^{\rm obs}$. The variance of the likelihood function $\sigma_{\tau_j}^2$ is taken from the measurement uncertainties, where we implicitly assume that measurements of $\tau$ and $\thetad$ are sufficiently approximated by Gaussian PDFs. An estimate of $\Ftilde_{\rm MW,h}$ can then be obtained by assuming a given halo density profile or constructing a PDF for $\DMMWh$, and adopting a value for $A_\tau$.

\section{The Milky Way Halo}
\indent In order to determine how the Milky Way halo (and in turn other galaxy haloes) contributes to the scattering of FRBs, we must constrain the scattering contribution of the Galactic disk. In the following sections, we first determine the amount of scattering that can occur in the thick disk of the Galaxy using the distribution of pulsar scattering measurements and DMs at high Galactic latitudes. We then compare the scattering measurements of FRB 121102 and FRB 180916 to the scattering expected from the Galactic disk using NE2001, and explain discrepancies between the scattering predictions of the NE2001 and YMW16 Galactic disk models. Finally, in Section~\ref{sec:halo}, we constrain the scattering contribution of the Galactic halo, followed by discussion of scattering constraints from pulsars in the Magellanic Clouds.

\subsection{Scattering from the Thick Disk}\label{sec:disk}
\begin{figure}
    \centering
    \includegraphics[width=0.47\textwidth]{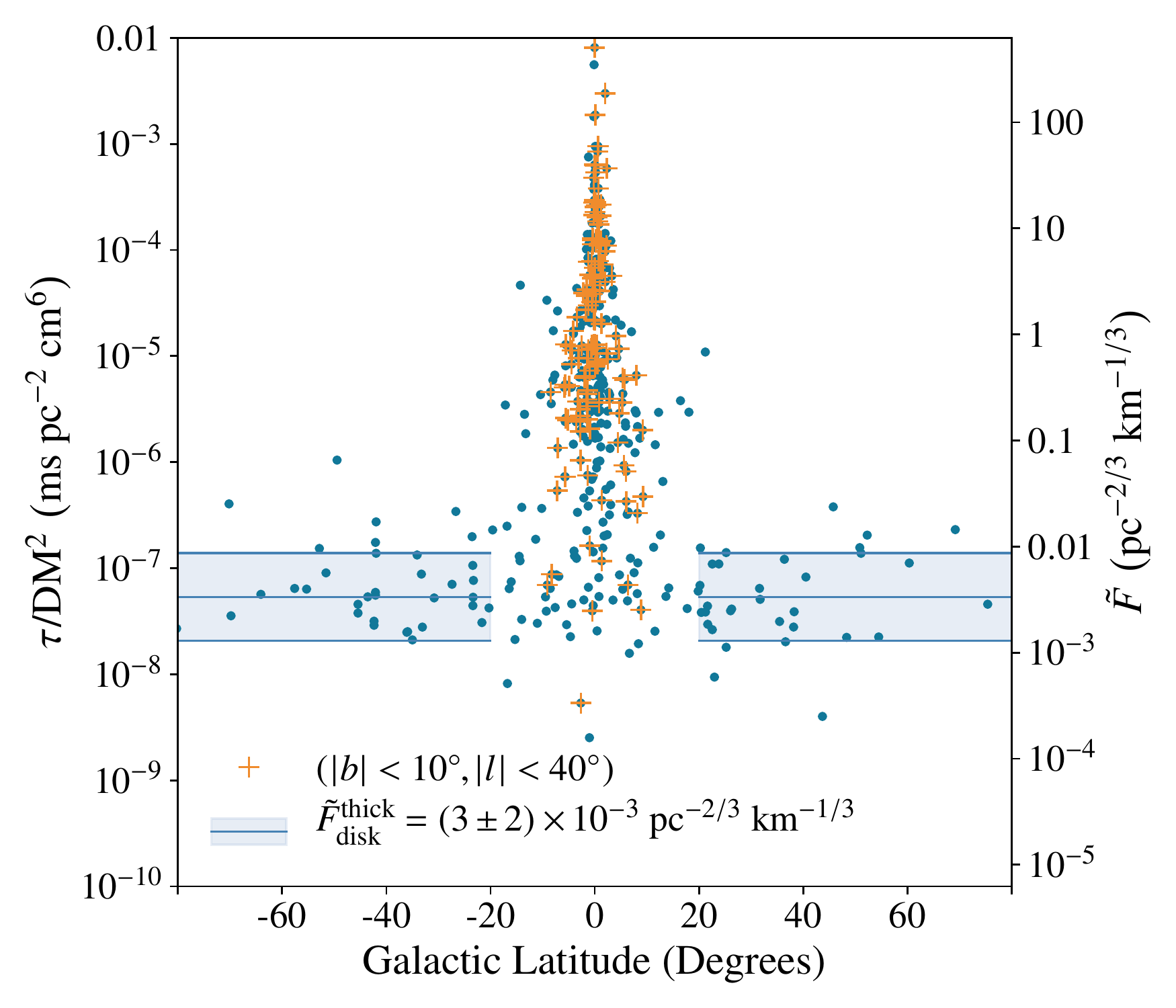}
    \caption{The distribution of $\tau/\DM^2$
    (which is directly proportional to $\Ftilde$)
    versus Galactic latitude for all Galactic pulsars, with $\tau$ in ms referenced to 1 GHz and DM in \DMunit. The average value and root-mean-square of the distribution for all pulsars above $\pm 20^\circ$ latitude is shown in blue. Pulsars closer to the Galactic center ($|b|<10^\circ$, $|l|<40^\circ$) are \replaced{highlighted in black}{shown as orange crosses}.}
    \label{fig:taudmb}
\end{figure}

Most currently known FRBs lie at high Galactic latitudes, and their LoS through the Galaxy predominantly sample the thick disk, which has a mean density at mid-plane of 0.015 cm$^{-3}$ and a scale height $\approx 1.6$ kpc \citep{2020ApJ...897..124O}. The distribution of $\tau/\DM^2$ for  Galactic pulsars with measurements of $\tau$
 \citep[][and references therein]{2016arXiv160505890C} and \DM\ and other parameters from 
\citet{2005AJ....129.1993M}\footnote{\url{http://www.atnf.csiro.au/research/pulsar/psrcat}} yields a direct constraint on $\Ftilde$: $\tau/\DM^2 \approx (16 \hspace{0.05in} {\rm ns})\Atau\Ftilde$, for $\nu = 1$ GHz and $G_{\rm scatt} = 1/3$ for sources embedded in the scattering medium. The distribution of $\Ftilde$ for all Galactic pulsars, assuming $\Atau\approx1$, is shown in Figure~\ref{fig:taudmb}. For the pulsars above $|20|^\circ$ Galactic latitude, the mean value of $\tau/\DM^2$ from a logarithmic fit is  $(5.3^{+5.0}_{-3.3})\times10^{-8}$ ms pc$^{-2}$ cm$^{6}$, which yields $\Ftilde \approx (3\pm2)\times10^{-3}$ \Funit. The value of $\Ftilde$ based on high-latitude pulsars is consistent with the related $F = l_i^{1/3} \Ftilde$ factor used in the NE2001 model for scattering in the thick disk.

\indent A structural enhancement to radio scattering for LoS below $|20|^\circ$ is reflected in the distribution of $\tau/\DM^2$ shown in Figure~\ref{fig:taudmb}, which shows a multiple orders of magnitude increase in $\Ftilde$ at low latitudes, with the largest values of $\Ftilde$ dominating LoS to the Galactic center. This latitudinal and longitudinal dependence of $\Ftilde$ is directly responsible for the ``hockey-stick" relation between $\tau$ and DM for Galactic pulsars, in which high-DM pulsars lying close to the Galactic plane and towards the Galactic center have a much steeper dependence on $\DM$ than pulsars lying high above the Galactic plane or towards the Galactic anti-center \citep[e.g.,][]{2004ApJ...605..759B, 2015ApJ...804...23K, 2016arXiv160505890C}. The implications of the directional dependence of $\Ftilde$ for FRB LoS are discussed further in Section~\ref{sec:ymw16}. 

\indent For the many FRBs in high Galactic latitude directions, the Galactic disk has a virtually undetectable contribution to the observed pulse broadening. The DM contribution of the thick disk is about ($23.5 \times {\rm csc}(|b|)$) \DMunit, which varies negligibly with longitude for $|b|>20^\circ$ \citep{2020ApJ...897..124O}. The pulse broadening at 1 GHz expected from the thick disk therefore ranges from $\tau < 0.25$ $\mu$s at $|b|=20^\circ$ to $\tau < 29$ ns at $|b| = 90^\circ$. As discussed in the following section, scattering from the Galactic thin disk and spiral arms increases dramatically for FRB LoS close to the Galactic plane. 

\subsection{Scattering Constraints for Two FRB Case Studies}\label{sec:frbs}

Unlike Galactic pulsars, for which the scintillation bandwidth and pulse broadening both result from the same electrons and conform to the uncertainty relation $2\pi\Delta\nu_{\rm d}\tau_{\rm d} = C_1$, 
some FRBs indicate that two \added{scattering} media are involved. \deleted{The scintillation bandwidth is caused by} \added{In these cases, the scintillation bandwidth is consistent with diffractive interstellar scintillation caused by} foreground Galactic turbulence while \added{the} pulse broadening \deleted{is extragalactic in origin, most likely from the host galaxy} \added{also has contributions from an extragalactic scattering medium \citep{ 2015Natur.528..523M, 2018ApJ...863....2G, 2019ARA&A..57..417C}}. Here we analyze the Galactic scintillations of two FRBs \added{with highly precise scattering measurements} in order to place constraints on any scattering in the Galactic halo. 

\subsubsection{FRB 121102}\label{sec:121102}
\begin{figure}
    \centering
    \includegraphics[width=0.47\textwidth]{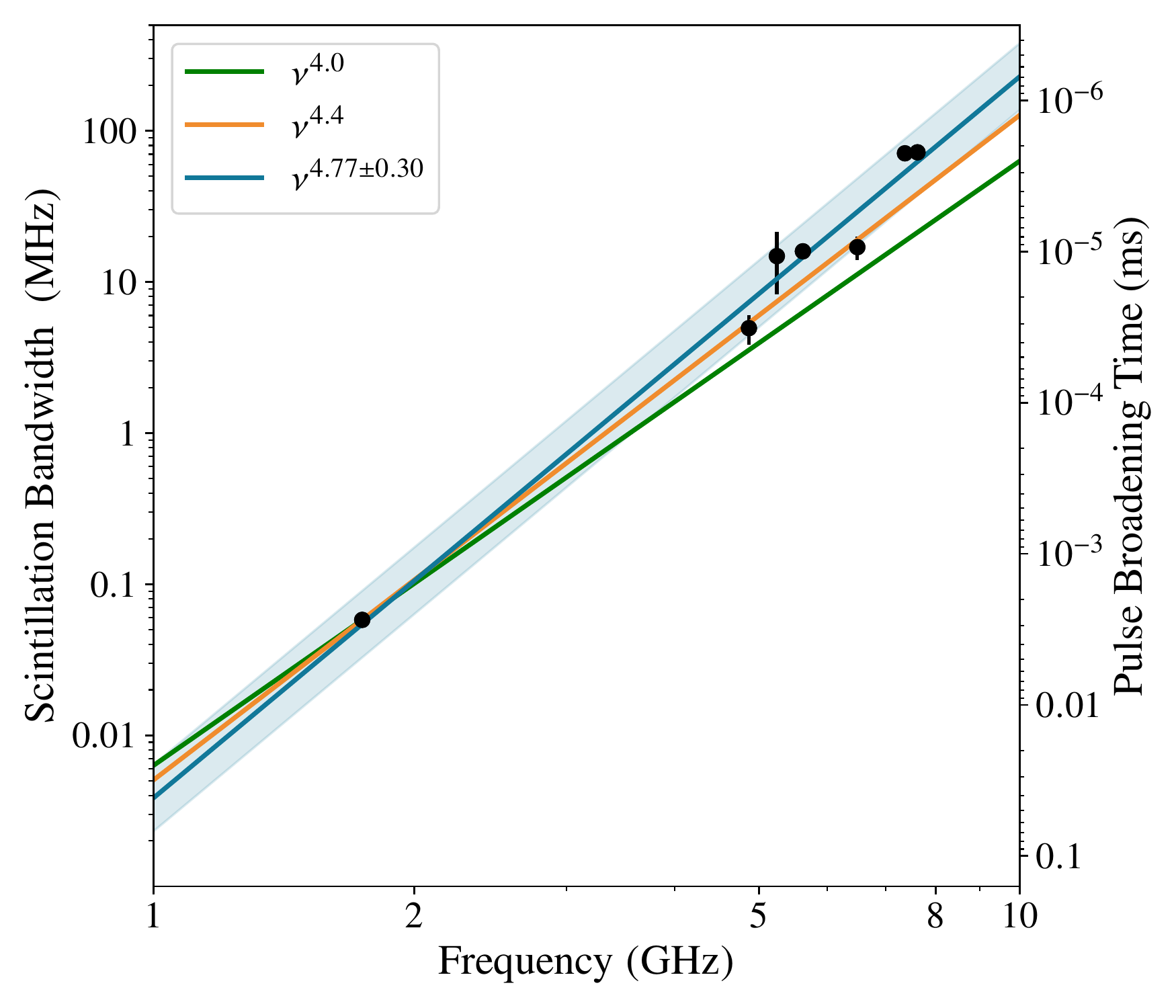}
    \caption{Scintillation bandwidth vs. radio frequency for FRB 121102. The  blue line and shaded region designates the least-squares fit and errors of ${\rm log}(\nud)$ vs. $\rm log(\nu)$. The green and orange lines are anchored to the lowest frequency point with the indicated frequency dependences. The corresponding pulse broadening time is shown on the right-hand axis assuming $\tau = 1/(2\pi\nud)$. Data points are from \cite{2019ApJ...876L..23H}, \cite{2018Natur.553..182M}, \cite{2018ApJ...863..150S}, and \cite{2018ApJ...863....2G}.}
    \label{fig:dissbw}
\end{figure}

FRB 121102 currently has the most comprehensive set of scattering constraints on an FRB source so far, with scintillation bandwidths $\nud = 58.1 \pm 2.3$ kHz at 1.65 GHz \citep{2019ApJ...876L..23H}, 5 MHz at 4.5 GHz \citep{2018Natur.553..182M}, $6.4\pm1.4$ MHz at 4.85 GHz \citep{2018ApJ...863..150S}, and $10-50$ MHz between 5 and 8 GHz \citep{2018ApJ...863....2G}. All of these scintillation bandwidth measurements are assembled in Figure~\ref{fig:dissbw}, along with a power-law fit to the data using linear least squares that gives a mean value $\nud = 3.8^{+2.5}_{-1.5}$ kHz at 1 GHz.

\indent FRB 121102 also has a pulse broadening time upper limit of $\tau < 9.6$ ms at 500 MHz \citep{2019ApJ...882L..18J}, and angular broadening $\thetad = 2\pm 1$ mas at 1.7 GHz and $\thetad \sim 0.4-0.5$ mas at 5 GHz, which was measured in the \cite{2017ApJ...834L...8M} high-resolution VLBI study of the FRB and its persistent radio counterpart. These angular diameters are consistent with those reported in \cite{2017Natur.541...58C} and with the NE2001 prediction. The scattering measurements are shown in Table~\ref{tab:measures} and are referenced to 1 GHz assuming a $\tau \propto \nu^{-4}$ frequency scaling. 

\begin{figure}
    \centering
    \includegraphics[width=0.5\textwidth]{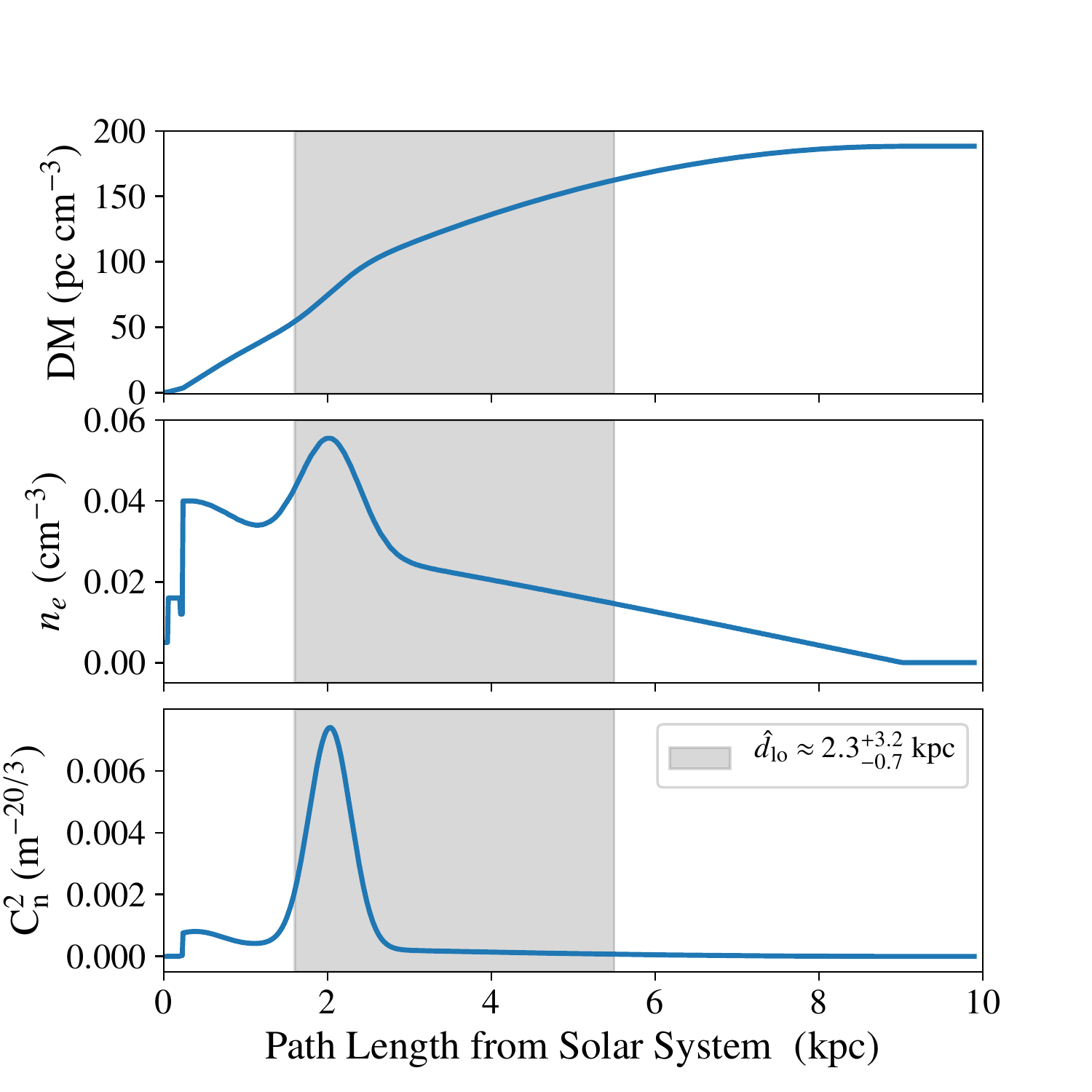}
    \caption{Galactic DM, electron density $n_e$, and $\cnsq$ contribution predicted by NE2001 for FRB 121102. The maximum DM (excluding the Galactic halo) is 188 \DMunit. The sharp changes in $n_e$ and $\cnsq$ between 0 and 0.3 kpc are due to  structure in the local ISM. The shaded grey region indicates the distance to the scattering screen derived from a numerical joint-probability analysis of the measured scattering constraints for FRB~121102; see Figure~\ref{fig:thetad}.}
    \label{fig:ne2001}
\end{figure}

\indent The NE2001 angular broadening and scintillation bandwidth predictions for FRB 121102 are broadly consistent with the corresponding empirical constraints (see Table~\ref{tab:measures}). In NE2001, the scattering for this LoS is dominated by an outer spiral arm located 2 kpc away and by the thick disk, which extends out to 17 kpc from the Galactic center \citep{2002astro.ph..7156C}. The $\cnsq$, electron density, and DM predicted by NE2001 along the LoS to FRB 121102 are shown in Figure~\ref{fig:ne2001}. Modeling of the anti-center direction in NE2001 is \added{independent of our analysis and is} based on DM and scattering measurements of \added{Galactic} pulsars in the same general direction and upper bounds on the angular scattering of extragalactic sources. The \added{NE2001} model parameters were constrained through a likelihood analysis of these measurements, which revealed that the fluctuation parameter was smaller for LoS that probe the outer Galaxy compared to the inner Galaxy \citep{1998ApJ...497..238L}. This result required that the thick disk component of NE2001 have a smaller fluctuation parameter compared to the thin disk and spiral arm components that are relevant to LoS through the inner Galaxy. 

\indent Since the measured $\nud$ and $\thetad$ for FRB 121102 are broadly consistent with the predicted amount of scattering from the Galactic disk, we use $\nud$ and $\thetad$ to estimate the effective distance to the dominant scattering material. For thin-screen scattering of a source located at a distance $\dso$ from the observer, the scattering diameter $\theta_{\rm s}$ is related to the observed angular broadening by
\begin{equation}
    \thetad \sim \theta_{\rm s} ( \dsl / \dso) = \theta_s \bigg(1 - \frac{\dlo}{\dso}\bigg),
\end{equation}
where $\dsl$ is the source-to-screen distance and $\dlo$ is the screen-to-observer distance. The scattering diameter is related to the pulse broadening delay by $\tau \approx \Atau \dso(\dsl/\dso)(1-\dsl/\dso)\theta_{\rm s}^2/8 ({\rm ln} 2)c$ \citep{2019ARA&A..57..417C}. For a thin screen near the observer and an extragalactic source, $\dlo \ll \dsl$, giving $\thetad \approx \theta_{\rm s}$ and
\begin{equation}\label{eq:ds}
    \thetad \approx \left(\frac{4({\rm ln}2) \Atau C_1 c}{ \pi \nud \dlo} \right)^{1/2}.
\end{equation}
The scattering screen location can thus be directly estimated from measurements of $\thetad$ and $\nud$.

\begin{figure}
    \centering
    \includegraphics[width=0.47\textwidth]{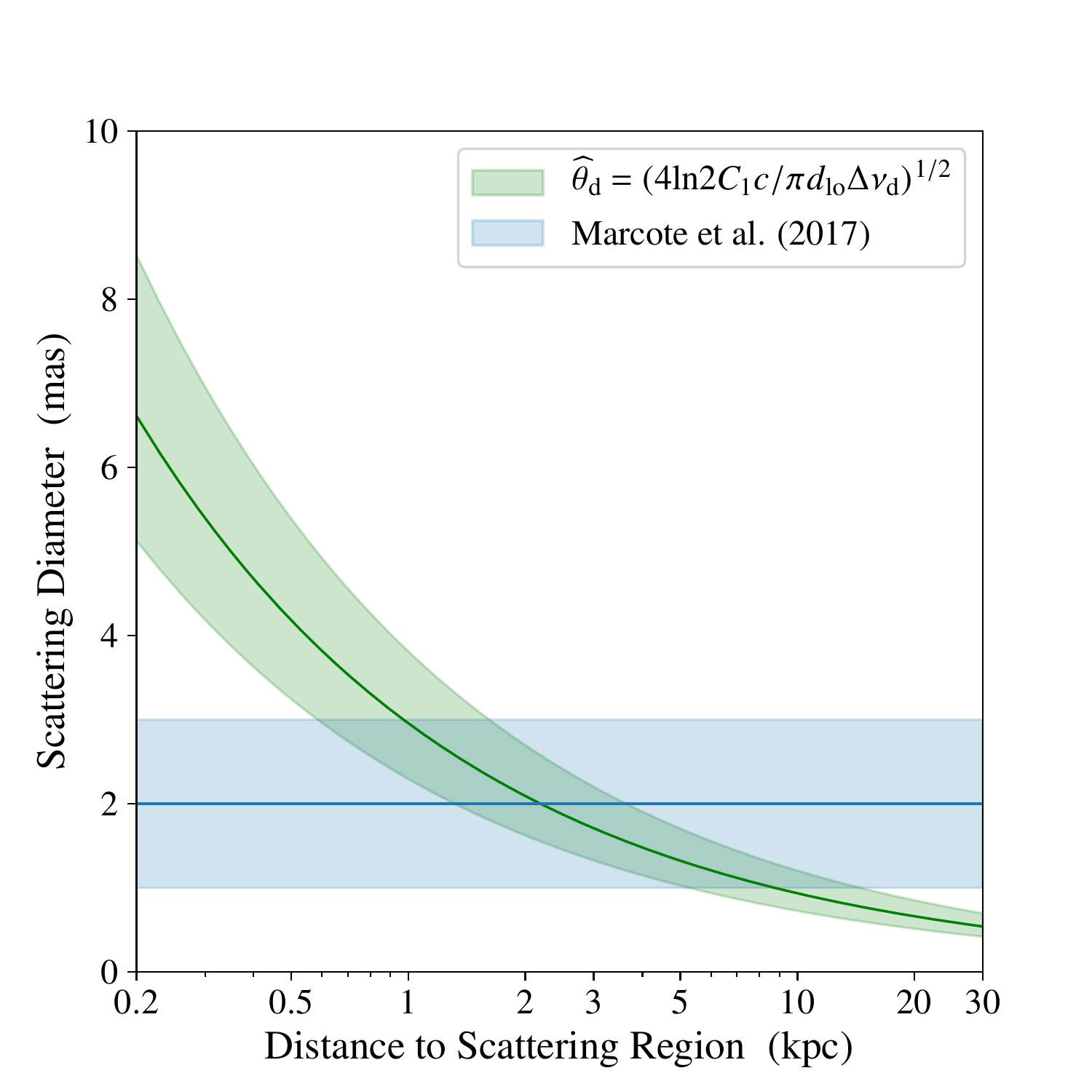}
    \caption{Scattering diameter vs. effective distance. The blue band indicates the angular diameter $\thetad = 2\pm1$ mas (1$\sigma$ errors) from \cite{2017ApJ...834L...8M}.  The green band is the predicted angular diameter for a thin-screen that matches the least-square fit to scintillation bandwidth measurements and the uncertainties at $\pm1\sigma$. All values are expressed at 1.7 GHz. A numerical joint-probability estimate constraining the overlap of the green and blue regions gives a screen distance $\hat{d}_{\rm lo} \approx 2.3^{+3.2}_{-0.7}$ kpc.
    }
    \label{fig:thetad}
\end{figure}

\indent Assuming that the same Galactic scattering material gives rise to both the angular broadening and the scintillation of FRB 121102, Equation~\ref{eq:ds} implies the scattering material has an effective distance $\hat{d}_{\rm lo} \approx 2.3$ kpc from the observer (assuming $\Atau \approx 1$), which is consistent with the distance to the spiral arm, as shown in Figure~\ref{fig:ne2001}. A numerical joint-probability analysis of the uncertainties in $\nud$ and $\thetad$, assuming both quantities follow normal distributions, allows the screen to be as close as 1.6 kpc or as far as 5.5 kpc (corresponding to the 15$\%$ and 85$\%$ confidence intervals). Figure~\ref{fig:thetad} shows a comparison of the relationship between $\thetad$ and $\dlo$ for a thin screen and the measured $\thetad$ from \cite{2017ApJ...834L...8M}. Given that there is no known HII region along the LoS to FRB~121102, the most likely effective screen is in fact the spiral arm.

\indent The scintillation bandwidth implies a pulse broadening contribution from the Milky Way disk and spiral arms $\tau_{\rm MW,d} \approx 0.04 \pm 0.02$ ms at 1 GHz. The upper limit on $\tau$ measured by \cite{2019ApJ...882L..18J} is an order of magnitude larger than the $\tau_{\rm MW,d}$ inferred from $\nud$. Any additional scattering beyond the Galactic contribution is more likely from the host galaxy due to the lack of intervening galaxies along the LoS, and the small amount of scattering expected from the IGM \citep{2013ApJ...776..125M, 2020arXiv201108519Z}.

\begin{deluxetable*}{c c c c c c }\label{tab:measures}
\tablecaption{DM and Scattering Referenced to 1 GHz of FRB 121102 and FRB 180916}
\tablewidth{0pt}
\tablehead{\multicolumn{6}{c}{Measurements} \\ \hline
\colhead{FRB}  & \colhead{DM (\DMunit)} & \colhead{$\tau$ (ms)$^{(1, 2)}$} & $\nud$ (kHz)$^{(3,4)}$ & \colhead{$\thetad$ (mas)$^{(5)}$} & \colhead{$\hat{d}_{\rm lo}$ (kpc)}}
\startdata
    121102 & $557$ & $<0.6$ & $3.8^{+2.5}_{-1.5}$ & $5.8 \pm 2.9$ & $2.3^{+3.2}_{-0.7}$ \\
        180916 & $349$ & $<0.026$ & $7.1\pm1.5$& \nodata & \nodata \\
    \hline \hline
    \multicolumn{6}{c}{Models}\\ \hline 
    FRB & DM (\DMunit) & $\tau$ (ms) & $\nud$ (kHz) & $\thetad$ (mas) & $\dlo$ (kpc) \\ \hline
    \multicolumn{6}{c}{NE2001}\\
    \hline 
    121102 & $188$ & $0.016$ & $11$ & $6$ & $2.05$ \\
    180916 & $199$ & $0.015$ & $12$ & $5$ & $2.5$\\
    \hline 
    \multicolumn{6}{c}{YMW16}\\
    \hline 
    121102 & $287$ & $0.84$ & $0.2$ & \nodata & \nodata \\
    180916 & $243$ & $0.42$ & $0.4$ & \nodata & \nodata \\
\enddata
\tablecomments{The top of the table shows the observed DM and scattering for FRB 121102 and FRB 180916. Since the scintillation and angular broadening measurements are broadly consistent with Galactic foreground predictions by NE2001, we emphasize that they are Galactic, whereas the pulse broadening may have an extragalactic contribution from the host galaxy. The scattering measurements are referenced to 1 GHz assuming $\tau \propto \nu^{-4}$ unless otherwise noted.\\ 
\indent The bottom of the table shows the asymptotic DM and scattering predictions of NE2001 and YMW16. NE2001 adopts $C_1 = 1.16$ to convert between $\nud$ and $\tau$, so we also use this value to calculate $\nud$ from $\tau$ for YMW16. YMW16 predictions were calculated in the IGM mode, but we only report the Galactic component of DM predicted by the model for comparison with NE2001.\\
\indent References: (1) \cite{2019ApJ...882L..18J}; (2) \cite{2020ApJ...896L..41C}; (3) this work (see Figure~\ref{fig:dissbw}; referenced to 1 GHz using a best-fit power law); (4) \cite{2020Natur.577..190M}; (5) \cite{2017ApJ...834L...8M}.}
\end{deluxetable*}

\subsubsection{FRB 180916}\label{sec:180916}
The scattering constraints for FRB 180916 consist of a scintillation bandwidth $\nud = 59 \pm 13$ kHz at 1.7 GHz \citep{2020Natur.577..190M} and a pulse broadening upper limit $\tau < 1.7$ ms at 350 MHz \citep{2020ApJ...896L..41C}. The $\nud$ and $\tau$ upper limit are entirely consistent with each other, so we again use $\nud$ and the inferred $\tau_{\rm MW,d}$ for the rest of the analysis due to its higher precision. Based on $\nud$, $\tau_{\rm MW,d} = 0.023\pm0.005$ ms at 1 GHz. As with FRB 121102 the NE2001 scattering predictions for this LoS are consistent with the empirical constraints to within the model's uncertainty, suggesting that the Galactic halo has a small ($\lesssim \mu$s level) contribution to the observed scattering.

\subsubsection{Comparison with YMW16 Scattering Predictions}\label{sec:ymw16}

The YMW16 model significantly overestimates the scattering of FRB 121102 and FRB 180916. The DM and scattering predictions for these FRBs are shown in Table~\ref{tab:measures}. Evaluating YMW16 for FRB 121102  using the \texttt{IGM} mode gives \replaced{$\log(\tau)=-3.012$}{$\log(\tau)=-3.074$} with $\tau$ in seconds, implying  $\tau = 0.84$~ms at 1~GHz, corresponding to $\nud \approx 0.2$~kHz, about 50 times smaller than the NE2001 value. Compared to the measured scattering, the nominal output of the YMW16 model overestimates the scattering \added{toward FRB 121102} by a factor of 28 to 35 (depending on whether a $\nu^{-4}$ or $\nu^{-4.4}$ scaling is used). Moreover, combining the measured $\thetad$ with the YMW16 estimate for $\tau$ implies a scattering screen distance $\sim 500$ kpc, beyond any local Galactic structure that could reasonably account for the scattering. For FRB 180916, YMW16 also overestimates $\tau$ to be 0.42~ms at 1~GHz, implying a scintillation bandwidth of 0.4~kHz at 1~GHz.
 
\indent The discrepancies between the observed scattering and the YMW16 predictions are due to several important factors. Unlike NE2001, YMW16 does not explicitly model electron density fluctuations.  Instead, it calculates DM for a given LoS and then uses the $\tau-\DM$ relation based on Galactic pulsars to predict $\tau$. In  using the $\tau-\DM$ relation, the YMW16 model incorrectly adjusts for the scattering of extragalactic bursts. The waves from extragalactic bursts are essentially planar when they reach the Galaxy, \deleted{which means that the Galactic plasma will scatter them more than it would the spherical waves from a Galactic pulsar at a distance of a few scale lengths of the electron density} \added{which means they are scattered from wider angles than diverging spherical waves from a Galactic pulsar would be}. The differences between plane and spherical wave scattering are discussed in detail for FRBs in \cite{2016arXiv160505890C}. The YMW16 model accounts for this difference by reducing the Galactic prediction of $\tau$ by a factor of two, when \added{geometric weighting of the mean-square scattering angle implies that} the \added{Galactic} scattering \added{prediction} should really be larger by a factor of three \added{to apply to extragalactic FRBs (see Eq.~10 in \citealt{2016arXiv160505890C})}.  This implies that values for $\tau$ in the output of YMW16 should be multiplied by a factor of six, which means that the model's overestimation of the scattering is really by  a factor of 170 to 208 when one only considers the correction for planar wave scattering.

\indent YMW16 may also overestimate the Galactic contribution to DMs of extragalactic sources viewed in the Galactic anti-center direction (and perhaps other low-latitude directions) because it significantly overestimates the observed DM distribution of Galactic pulsars that NE2001 is based on. In YMW16, the dominant DM contributions to extragalactic sources in this direction are from the thick disk and from the spiral arms exterior to the solar circle. Together these yield DM values of 287 \DMunit\ and 243 \DMunit\ for FRB~121102 and FRB~180916, respectively. These DM predictions are over $50\%$ and $20\%$ larger for each FRB than the NE2001 values, which may be due to overestimation of the densities or characteristic length scales of the outer spiral arm and thick disk components. 
 
\indent The primary cause for YMW16's scattering over-prediction is that the part of the pulsar-derived $\tau-\DM$ relation that applies to large values of DM should not be used for directions toward the Galactic anti-center. The $\tau-\DM$ relation has the empirical form \citep{2016arXiv160505890C}
\begin{equation}
\tau = (2.98\times10^{-7}\ {\rm ms})\, \DM^{1.4}(1+3.55\times10^{-5}\DM^{3.1})
\label{eq:taud_vs_DM}
\end{equation}
based on a fit to pulsar scattering data available through 2016. Similar fits were previously done by \cite{1997MNRAS.290..260R}, \cite{2004ApJ...605..759B},  and \cite{2015ApJ...804...23K}. It scales as $\DM^{1.4}$ for $\DM \lesssim 30$ \DMunit and as $\DM^{4.5}$ for $\DM \gtrsim 100$ \DMunit.    

\indent The YMW16 model uses Krishnakumar's model for 327~MHz scattering times scaled to 1~GHz by a factor $(0.327)^4$, giving
$\tau = (4.1\times10^{-8}\ {\rm ms})\ \DM^{2.2}(1+0.00194\times\DM^{2})$. This scaling law is in reasonable agreement with the expression in Equation~\ref{eq:taud_vs_DM} except at very low DM values. The Krishnakumar scaling law adopted the \cite{1997MNRAS.290..260R} approach of fixing the leading DM exponent to 2.2, which is based on the assumption that the relatively local ISM is uniform. This assumption is imperfect given our knowledge of the Local Bubble and other shells and voids in the local ISM. The steep $\tau\propto \DM^{4.2 \ {\rm to }\ 4.5}$ scaling for Galactic pulsars is from LoS that probe the inner Galaxy, where the larger star formation rate leads to a higher supernova rate that evidently affects the turbulence in the HII gas, and results in a larger $\Ftilde$, as shown in Section~\ref{sec:disk}.

\indent If the YMW16 model were to use instead the shallow part of the $\tau-\DM$ relation,  $\tau\propto \DM^{2.2 \ {\rm to} \ 1.4}$, which is more typical of LoS through the outer Galaxy,  its scattering time estimates would be smaller by a  factor of $\sim 1.94\times 10^{-3} \times (287)^{2} \sim 160$ or $\sim 3.55\times 10^{-5} \times (287)^{3.1} \sim 1500$, depending on which $\tau-\DM$ relation is used (and using FRB 121102 as an example). These overestimation factors could be considerably smaller if smaller DM values were used. While there is a considerable range of values for the overestimation factor based on the uncertainties in the empirical scaling law, it is reasonable to conclude that the high-DM part of the $\tau-\DM$ relation should not be used for the anti-center FRBs.

\subsection{Fluctuation Parameter of the Galactic Halo}\label{sec:halo}
The measurements of $\tau$ and $\thetad$ for FRBs 121102 and 180916, combined with the scattering predictions of NE2001, yield a maximum likelihood estimate for the pulse broadening contribution of the Milky Way halo (see Equation~\ref{eq:Like}). \deleted{The likelihood function is shown in Figure~6. The likelihood function is positive for values of $(\Ftilde \times \DM^2)_{\rm MW,h}$ extending to zero because the measured scattering from both FRBs is close enough to the  NE2001 predictions that the halo could have a negligible scattering contribution.} The $95\%$ upper confidence interval yields $(\Ftilde \times \DM^2)_{\rm MW,h} < 250/A_\tau$ \FDMunit. The maximum amount of pulse broadening expected from the Galactic halo is therefore $\tau_{\rm MW,h} < 12$ $\mu$s at 1 GHz, which is comparable to the scattering expected from the Galactic disk for LoS towards the Galactic anti-center or at higher Galactic latitudes.

\indent Based on \deleted{a meta-analysis of} the broad range of $\DMMWh$ currently consistent with the empirical and modeled constraints (see Section~\ref{sec:halomodels}), we construct a Gaussian probability density function (PDF) for $\widehat{\DM}_{\rm MW,h}$ with a mean of 60 \DMunit\ and $\sigma_{\rm \widehat{DM}} = 18$ \DMunit \deleted{, which is shown in Figure~6}. \added{Combining} this PDF \added{with the maximum likelihood estimate for $(\Ftilde \times \DM^2)_{\rm MW,h}$} yields an upper limit $\Ftilde_{\rm MW,h} < 0.03/A_\tau$ \Funit. While $A_\tau$ is probably about 1, if $A_\tau$ is as small as $1/6$ then $\Ftilde_{\rm MW,h}$ could be up to 6 times larger. 

\indent \added{This estimate of $\Ftilde_{\rm MW,h}$ is based on just two LoS towards the Galactic anti-center, and it is unclear how much $\Ftilde_{\rm MW,h}$ will vary between different LoS through the halo. Given that sources viewed through the inner Galaxy (near $b = 0^\circ$) are more heavily scattered, it is unlikely that estimates of $\Ftilde_{\rm MW,h}$ will be obtainable for LoS that intersect the halo through the inner Galaxy. However, FRB 121102, FRB 180916, and most of the FRBs detected at higher latitudes do not show evidence of any intense scattering regions that might be associated with the halo \citep[e.g.,][]{2020MNRAS.497.1382Q}, suggesting that extremely scattered FRBs would be outliers and not representative of the Galactic halo's large-scale properties. All FRBs are ultimately viewed through not only the Galactic halo but also the haloes of their host galaxies and, in some cases, the haloes of intervening galaxies. Given the observed variations in scattering between different LoS through the Milky Way, it appears most likely that the heaviest scattered FRBs will be viewed through scattering regions within galaxy disks rather than haloes, and extrapolating our analysis to a larger sample of FRBs will require determining whether observed variations in $\Ftilde$ are due to variations between galaxy haloes, disks, or the sources' local environments.} In the following sections, we compare the Milky Way halo scattering contribution \added{inferred from FRBs 121102 and 180916} to scattering observed from the Magellanic Clouds and galaxy haloes intervening LoS to FRBs.

\subsection{Constraints from Pulsars in the Magellanic Clouds}\label{sec:MCs}
At distances of 50 to 60 kpc and latitudes around $-30^\circ$, pulsar radio emission from the Large and Small Magellanic Clouds (LMC/SMC) mostly samples the Galactic thick disk and a much smaller path length through the Galactic halo than FRBs. So far, twenty-three radio pulsars have been found in the LMC and seven in the SMC \citep[e.g.,][]{1991MNRAS.249..654M,2001ApJ...553..367C,2006ApJ...649..235M, 2013MNRAS.433..138R, 2019MNRAS.487.4332T}. Very few scattering measurements exist for these LoS. PSR B0540$-$69 in the LMC has a DM of 146.5 \DMunit\ and was measured to have a pulse broadening time $\tau = 0.4$ ms at 1.4 GHz \citep{2003ApJ...590L..95J}. The Galactic contribution to DM and scattering predicted by NE2001 towards this pulsar are $\DM_{\rm NE2001} = 55$ \DMunit, and $\tau_{\rm NE2001} = 0.3\times10^{-3}$ ms at 1 GHz. Based on this DM estimate, the pulsar DM receives a contribution of about $92$ \DMunit\ from the LMC and the Galactic halo. The lowest DMs of pulsars in the LMC have been used to estimate the DM contribution of the halo to be about 15 \DMunit\ for pulsars in the LMC \citep{2020ApJ...888..105Y}, which suggests that the LMC contributes about 77 \DMunit\ to the DM of B0540$-$69. 

\indent The scattering observed towards B0540$-$69 is far in excess of the predicted scattering from the Galactic disk. Since B0540$-$69 not only lies within the LMC but also within a supernova remnant, it is reasonable to assume that most of the scattering is contributed by material within the LMC. Using the upper limit $\Ftilde_{\rm MW,h} < 0.03/A_\tau$ \Funit\ and $\widehat{\DM}_{\rm MW,h} = 15$ \DMunit\ yields $\tau_{\rm MW,h} < 0.1$ $\mu$s at 1.4~GHz for this LoS, which is too small to explain the observed scattering. If we instead combine $\tau = 0.4$ ms at 1.4~GHz and the estimated $\widehat{\DM}_{\rm LMC} = 77$ \DMunit, \added{we find $\Ftilde_{\rm LMC} \approx 16/A_\tau$} \deleted{$\Ftilde_{\rm LMC} \approx 4.2/A_\tau$} \Funit. More scattering measurements for LMC and SMC pulsars are needed to better constrain the fluctuation parameters of the LMC and SMC, which in turn will improve our understanding of interstellar plasma in these satellite galaxies.

\section{Constraints on Intervening Haloes along Lines of Sight to FRBs}\label{sec:conc}
As of this paper, two FRBs are found to pass through galactic haloes other than those of their host galaxies and the Milky Way: FRB 181112, which passes within 30 kpc of the galaxy DES J214923.89$-$525810.43, otherwise known as FG$-$181112 \citep{2019Sci...366..231P}, and FRB 191108, which passes about 18 kpc from the center of M33 and 185 kpc from M31 \citep{2020MNRAS.tmp.2810C}. Both FRBs have measurements of $\tau$ which are somewhat constraining. 

\subsection{FRB 181112}

\indent FRB 181112 was initially found to have $\tau < 40$ $\mu$s at 1.3 GHz by \cite{2019Sci...366..231P}; follow-up analysis of the ASKAP filterbank data and higher time resolution data for this burst yielded independent estimates of $\tau < 0.55$ ms \citep{2020MNRAS.497.1382Q} and $\tau \approx 21 \pm 1$ $\mu$s \citep{2020ApJ...891L..38C} at 1.3 GHz. We adopt the last value for our analysis, with the caveat that the authors report skepticism that the data is best fit by a pulse broadening tail following the usual frequency dependence expected from scattering in a cold plasma, and that the measured decorrelation bandwidth of the burst spectrum is in tension with the pulse broadening fit (for a full discussion, see Section 4 of \citealt{2020ApJ...891L..38C}).

\indent We first place an upper limit on $(\Ftilde \times \DM^2)_{\rm i,h}$ for FG$-$181112 by assuming that the intervening halo may contribute up to all of the observed scattering of FRB 181112. The halo density profile (Equation~\ref{eq:neh}) is re-scaled to the lens redshift ($z_{\rm i,h} = 0.36$) and evaluated at the impact parameter $R_\perp = 29$ kpc. \cite{2019Sci...366..231P} constrain the mass of the intervening halo to be $M_{\rm halo}^{\rm FG-181112} \approx 10^{12.3}M_\odot$. Again assuming a physical extent to the halo of $2r_{200}$ gives \deleted{an effective} \added{a} path length through the halo $L \approx 930$ kpc. The observed scattering $\tau \approx 21$ $\mu$s is the maximum amount of scattering that could be contributed by the halo, i.e., $\tau_{\rm i,h} < 21$ $\mu$s at 1.3 GHz, which yields $(\Ftilde \times \DM^2)_{\rm i,h} < 13/A_\tau$ \FDMunit\ using Equation~\ref{eq:taudmnuz}.

\indent Assuming the halo density profile where $y_0 = \alpha = 2$ gives $\widehat{\DM}_{\rm i,h} \approx 135$ \DMunit\ in the frame of the intervening galaxy. This DM estimate for the halo is similar to the estimate of $122$ \DMunit\ from \cite{2019Sci...366..231P}, but as they note, the DM contribution is highly sensitive to the assumed density profile and could be significantly smaller if the physical extent and/or the baryonic fraction of the halo are smaller. This DM estimate yields $\Ftilde_{\rm i,h} < (7\times10^{-4})/A_\tau$ \Funit. If the DM contribution of the intervening halo is smaller, then $\Ftilde_{\rm i,h}$ could be up to an order of magnitude larger. The total observed DM of the FRB is broadly consistent with the estimated DM contributions of the Milky Way, host galaxy, and IGM alone \citep{2019Sci...366..231P}, so the uncertainty in $\DM_{\rm i,h}$ remains the greatest source of uncertainty in deconstructing $(\Ftilde \times \DM^2)_{\rm i,h}$. Both estimates of $(\Ftilde \times \DM^2)_{\rm i,h}$ and $\Ftilde_{\rm i,h}$ for FRB 181112 are within the upper limits for the Milky Way halo. 

\subsection{FRB 191108}

\indent FRB 191108 passes through both the M31 and M33 haloes and has a source redshift upper limit $z\lesssim0.5$ based on DM. \cite{2020MNRAS.tmp.2810C} report an upper limit of $80$ $\mu$s on the intrinsic pulse width and scattering time at 1.37 GHz, but they demonstrate that this limit is likely biased by dispersion smearing. \cite{2020MNRAS.tmp.2810C} also report $25\%$ intensity modulations at a decorrelation bandwidth $\sim 40$ MHz. This decorrelation bandwidth may be attributable to scattering in the M33 halo and/or in the host galaxy (for a full discussion, see Section 3.4 of \citealt{2020MNRAS.tmp.2810C}).

\indent Re-scaling our galactic halo density profile using halo masses $M_{\rm halo}^{\rm M33} \approx 5\times10^{11}M_\odot$ and $M_{\rm halo}^{\rm M31} \approx 1.5\times10^{12}M_\odot$ yields a total DM contribution from both haloes of about 110 \DMunit, nearly two times larger than the DM contribution estimated by \cite{2020MNRAS.tmp.2810C}, who use a generic model for the M33 and M31 haloes from \cite{2019MNRAS.485..648P} based on the same galaxy masses. We assume that the density profiles are independent; if there are dynamical interactions between the haloes then these may slightly modify the overall density distribution along the LoS, but it is unclear how turbulence in the plasma would be affected, if at all. Since the impact parameter of 18 kpc for M33 is significantly smaller than the 185 kpc for M31, M33 dominates the predicted DM contribution to FRB 191108 (with $\widehat{\DM}_{\rm i,h} \approx 90$ \DMunit), and therefore is more likely than M31 to also dominate the scattering. 

\indent If we were to assume that $\nud\approx 40$ MHz (at 1.37 GHz, which translates to $\tau \approx 4$ ns) is attributable to scattering in the M33 halo, then we get $(\Ftilde \times \DM^2)_{\rm i,h} \approx 0.23/A_\tau$ \FDMunit\ for $z_{\rm host} = 0.5$. A smaller source redshift would increase $\dsl \dlo/\dso$, resulting in an even smaller value of $(\Ftilde \times \DM^2)_{\rm i,h}$. For a halo DM contribution of about 90 \DMunit\, this estimate of $(\Ftilde \times \DM^2)_{\rm i,h}$ yields $\Ftilde_{\rm i,h} \approx (2.7\times10^{-5})/A_\tau$ \Funit, which is three orders of magnitude smaller than the upper limit we infer for the Milky Way halo. Using a smaller $\DM_{\rm i,h} \approx 50$ \DMunit\ increases $\Ftilde$ to $\Ftilde_{\rm i,h} \approx (9\times10^{-5})/A_\tau$ \Funit. Generally speaking, if $\Ftilde$ is about a factor of 10 larger, then the pulse broadening from the halo would be 10 times larger and the scintillation bandwidth 10 times smaller. On the other hand, if M31 were to contribute more significantly to the DM then $\Ftilde_{\rm i,h}$ would be smaller than our estimate. While there is a range of reasonable values for $\Ftilde_{\rm i,h}$, it appears that scattering in the M33 halo is negligible.

\indent \cite{2020MNRAS.tmp.2810C} use a different approach to evaluate the scattering of FRB 191108. They estimate a scattering angle from the decorrelation bandwidth in order to obtain an estimate of the diffractive scale and rms electron density fluctuations in the halo. Making assumptions about the outer scale and the relationship between the mean density and rms density fluctuations, they find a mean electron density for the halo that is larger than expected, and conclude that if the scattering occurs in M33, then it is more likely from cool clumps of gas embedded in the hot, extended halo. \cite{2019Sci...366..231P} use a similar methodology to estimate a mean density for the halo of FG$-$181112. Rather than make an indirect estimate of $n_e$ in each halo, our analysis yields a direct constraint on $(\Ftilde \times \DM^2)_{\rm i,h}$ from observable quantities. The corresponding estimates of $\Ftilde$ are sufficient to demonstrate that very little scattering occurs \added{along either of these FRB LoS through the} galaxy haloes. Further deconstructing $\epsilon^2$, $\zeta$, and $f$ from $\Ftilde$ will require more information about the outer and inner scales of turbulence, which may differ from halo to halo. 

\begin{figure}
    \centering
    \includegraphics[width=0.47\textwidth]{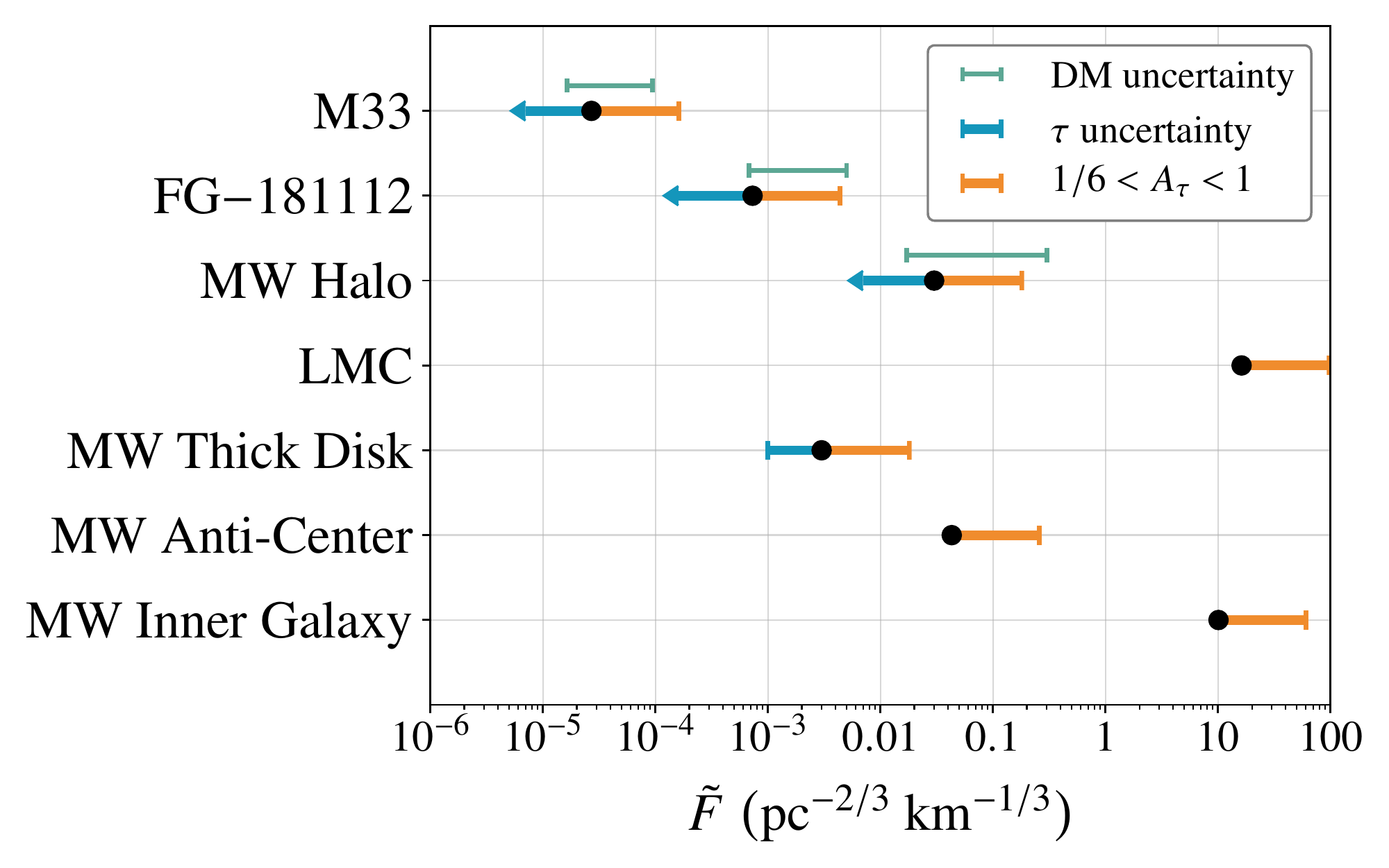}
    \caption{Nominal values of $\Ftilde$ for different components of the Milky Way (MW), the Large Magellanic Cloud (LMC), and for the foreground galaxies of FRB 181112 and FRB 191108. The Galactic anti-center and inner galaxy values are calculated by integrating NE2001 through the entire disk in the directions $(l = 130^\circ, b = 0^\circ)$ and $(l = 30^\circ, b=0^\circ)$, respectively. The nominal $\Ftilde$ for each halo was calculated assuming the modeled halo DMs discussed in the text. The orange error bars indicate the maximum values of $\Ftilde$ for $A_\tau > 1/6$, with $A_\tau = 1$ for the black points. The green error bars indicate a range of $\Ftilde$ for a representative range of halo DMs (and $A_\tau = 1$). The green lower and upper bounds correspond to: $20<\DM_{\rm MW,h}<120$ \DMunit\ for the Galactic halo, $50<\DM_{\rm i,h}<140$ \DMunit\ for FG$-$181112, and $50<\DM_{\rm i,h}<120$ \DMunit\ for M33. For each halo, the blue bar indicates that the scattering constraints are upper limits. The blue bar for the thick disk indicates the root-mean-square error in the distribution of $\Ftilde$ for high Galactic latitude pulsars.}
    \label{fig:Fsummaryplot}
\end{figure}

\section{Discussion}
We present a straightforward methodology for constraining the internal electron density fluctuations of galaxy haloes using FRB scattering measurements. The pulse broadening time $\tau \propto \Ftilde \times \DM^2$, where the fluctuation parameter $\Ftilde$ quantifies the amount of scattering per unit DM and is directly related to the density fluctuation statistics. We analyze two case studies, FRB 121102 and FRB 180916, and find their scattering measurements to be largely consistent with the predicted scattering from the Galactic disk and spiral arms, plus a small or negligible contribution from the Galactic halo. A likelihood analysis of their scintillation bandwidths and angular broadening places an upper limit on the product of the Galactic halo DM and fluctuation parameter $(\Ftilde \times \DM^2)_{\rm MW,h} < 250/A_\tau$ \FDMunit, where $A_\tau$ is the dimensionless constant relating the mean scattering time to the $1/e$ time of a scattered pulse. This estimate can be used to calculate the pulse broadening delay induced by electron density fluctuations in the halo, independent of any assumptions about the electron density distribution of the Galactic halo. The upper limit on $(\Ftilde \times \DM^2)_{\rm MW,h}$ implies a maximum amount of pulse broadening from the Galactic halo $\tau_{\rm MW,h} < 12$ $\mu$s at 1 GHz. 

\indent While the DM contribution of the Milky Way halo to FRB DMs is still poorly constrained, we adopt a Gaussian PDF for the observed DM of the halo to estimate $\Ftilde_{\rm MW,h} < 0.03/A_\tau$ \Funit. We compare this to the fluctuation parameter of the Galactic thick disk using the distribution of $\tau/\DM^2$ for all Galactic pulsars at high Galactic latitudes with pulse broadening measurements. We measure the fluctuation parameter of the thick disk to be $\Ftilde_{\rm disk}^{\rm thick} = (3\pm2)\times10^{-3}$ \Funit, about an order of magnitude smaller than the halo upper limit. At high Galactic latitudes, the thick disk will only cause a scattering delay on the order of tens of nanoseconds at 1 GHz. Larger samples of FRBs and continued X-ray observations of the Galactic halo will refine our understanding of the DM contribution of the halo and may modify our current constraint on $\Ftilde_{\rm MW,h}$\added{, which is only based on two LoS through the halo}. While we assume for simplicity that the density distribution of the halo is spherically symmetric, $\Ftilde_{\rm MW,h}$ and $\DMMWh$ will vary between different LoS through the halo, and an extension of our analysis to a larger sample of FRBs may yield a more constraining limit on the average fluctuation parameter of the halo.

\indent Extrapolating the scattering formalism we use for the Galactic halo to intervening galaxies, we examine two examples of FRBs propagating through intervening haloes, FRB 181112 and FRB 191108. The observed upper limits on each halo's contribution to $\tau$ are $\tau_{\rm i,h}<21$ $\mu$s at 1.3 GHz for FRB 181112 \citep{2020ApJ...891L..38C} and $\tau_{\rm i,h} < 4$ ns at 1.37 GHz for FRB 191108 \citep{2020MNRAS.tmp.2810C}. We find $(\Ftilde \times \DM^2)_{\rm i,h} < 13/A_\tau$ \FDMunit\ for FRB 181112 and $(\Ftilde \times \DM^2)_{\rm i,h} < 0.2/A_\tau$ \FDMunit\ for FRB 191108. Both estimates fall within the upper limit for the Milky Way halo, and all of these haloes have small to negligible scattering contributions \added{for the FRBs considered in this paper}.

\indent We also model the DM contribution of each intervening halo to find nominal constraints on $\Ftilde$. The values of $\Ftilde$ from our analysis of FRB 181112, FRB 191108, the LMC, the Galactic halo, the Galactic thick disk, and the values of $\Ftilde$ used in NE2001 for the Galactic anti-center and inner Galaxy are all assembled in Figure~\ref{fig:Fsummaryplot}. The uncertainties associated with the conversion factor $A_\tau$ and the halo DMs are also shown. \added{The values of $\Ftilde$ for M33, FG$-$181112, the Galactic halo, and the LMC are essentially point estimates because they are based on individual sources, while the estimates provided for the Galactic thick disk, anti-center, and inner Galaxy are based on the population of Galactic pulsars.} Broadly speaking, the $\Ftilde$ upper limit for the Galactic halo is similar to that of the disk and spiral arms in the anti-center direction, and is about an order of magnitude larger than the fluctuation parameter of the thick disk. The value of $\Ftilde$ for the LMC is similar to that of the inner Milky Way because it is based on the pulse broadening of B0540$-$69, which lies within a supernova remnant and hence within an enhanced scattering region. Our estimates of $\Ftilde_{\rm i,h}$ for both FRB 181112 and FRB 191108 indicate that very little scattering occurs in the haloes intervening their LoS. 

\indent The fluctuation parameter is directly related to the inner and outer scales of turbulence as $\Ftilde \propto (\zeta \epsilon^2/f)(\louter^2\linner)^{-1/3}$, where $\zeta$ and $\epsilon$ respectively describe changes in the mean density between different gas cloudlets and the variance of the density fluctuations within cloudlets. While the inner and outer scales in the Galactic warm ionized medium (WIM) are constrained by pulsar measurements to be on the order of \replaced{$\linner \sim 1000$}{$100 \lesssim \linner \lesssim 1000$} km and $\louter \gtrsim 10$ pc
\replaced{\citep{1995ApJ...443..209A}}{\citep{1990ApJ...353L..29S,1995ApJ...443..209A,2004ApJ...605..759B,2009MNRAS.395.1391R}}
, the corresponding scales in hot halo gas are probably much larger. Given the size of the halo, $\louter$ could be on the order of tens of kpc. The inner scale could also be larger if it is related to the proton gyroradius and the magnetic field strength is smaller in the halo than in the disk, which is probably the case given that the rotation measures of extragalactic sources tend to be larger closer to the Galactic plane \citep{2017ARA&A..55..111H}. Given that $f$, $\louter$, and $\linner$ are all probably larger in the halo than in the disk, we would expect $\Ftilde_{\rm MW,h}$ to be much smaller than $\Ftilde_{\rm disk}^{\rm thick}$. If we further expect the Milky Way halo to be similar to other galaxy haloes like FG$-$181112 and M33, then $\Ftilde_{\rm MW,h}$ would likely be less than $10^{-3}$ \Funit. However, our current constraints allow $\Ftilde$ to be larger in the halo than in the disk, which suggests that the upper limit for $\Ftilde_{\rm MW,h}$ is not constraining enough to make any further conclusions about $\zeta$, $\epsilon^2$, $f$, $\louter$, and $\linner$ in the halo.

\indent \added{On the other hand, quasar absorption studies of the CGM of other galaxies (mostly at redshifts $z\gtrsim2$) indicate the presence of $\sim10^4$ K gas \citep{2015Sci...348..779H, 2016ApJS..226...25L, 2018MNRAS.473.5407M}, suggesting that the CGM is a two-phase medium consisting of warm gas clumps embedded in a hot ($10^6$ K) medium \citep{2018MNRAS.473.5407M}. Using a cloudlet model based on the simulations of \cite{2018MNRAS.473.5407M}, \cite{2019MNRAS.483..971V} demonstrate that a clumpy CGM could significantly scatter FRBs. Our empirical constraints on $\Ftilde$ are largely independent of any assumptions about the physical properties of the scattering medium. We assume a halo density model to estimate the DM contribution of a halo, although mapping of ionized and neutral high-velocity clouds in the Galactic CGM indicate that the DM is likely dominated by the hot gas \citep{2019MNRAS.485..648P}. As a composite parameter, $\Ftilde$ is insensitive to a broad range of assumptions about gas temperature or clumps, and could serve as an independent test of the two-phase model for the CGM. In a clumpy, cooler CGM, the inner and outer scales of turbulence would be similar to those in the WIM and $f \ll 1$, and $\Ftilde$ would be larger than it would be in a hot medium with a larger filling factor and scale size. Adopting fiducial values of $\epsilon^2 = \zeta = 1$, $f\sim10^{-4}$ (the value used by \cite{2019MNRAS.483..971V}), $\linner \sim 100$ km, and $\louter \sim 10$ pc gives $\Ftilde \sim 500$ \Funit. This estimate is orders of magnitude larger than our results for the Galactic halo and the foreground haloes of FRBs 181112 and 191108, suggesting that halo gas probed by these LoS is either not dominated by cooler clumps, or that $f$, $\louter$, and $\linner$ are significantly different in the clumpy CGM than otherwise assumed by \cite{2018MNRAS.473.5407M} and \cite{2019MNRAS.483..971V}. }

\indent \deleted{A more stringent comparison of the hot halo and thick disk will require a stricter constraint on $\Ftilde$ for the halo.} \added{A more stringent comparison of hot gas in the halo and the WIM will require a larger sample of precise FRB scattering measurements.} Regardless, the nominal range of $\Ftilde$ constrained for the Galactic halo and the haloes intervening FRB 181112 and FRB 191108 demonstrate the range of internal properties that different galaxy haloes can have. A broader sample of FRB scattering measurements with intervening halo associations will expand this range and may potentially reveal an interesting diversity of galaxy haloes. 

\indent Many more FRBs with intervening galaxy haloes will likely be discovered in the near future. In these cases, the amount of scattering to be expected from the intervening haloes will depend not only on the fluctuation parameter $\Ftilde$ and DM of the halo, but also on the relative distances between the source, halo, and observer, and the effective path length through the halo. Depending on the relative configuration, an intervening halo may amplify the amount of scattering an FRB experiences by factors of 100 or more relative to the amount of scattering expected from the Milky Way halo. However, plausibly attributing scattering to an intervening halo will still require careful consideration of the FRB host galaxy, which in many cases may be the dominant source of FRB scattering.

\acknowledgements

The authors thank the referee for their useful comments and acknowledge support from the National Aeronautics and Space Administration (NASA 80NSSC20K0784) and the National Science Foundation (NSF AAG-1815242), and are members of the NANOGrav Physics Frontiers Center, which is supported by the NSF award PHY-1430284.

\bibliography{bib}

\listofchanges

\end{document}